\documentclass[aps,amsmath,amssymb,superscriptaddress,reprint]{revtex4-1}
\usepackage{dcolumn}% Align table columns on decimal point
\usepackage{bm}% bold math
\usepackage{amsmath}
\usepackage{amssymb}
\usepackage{amsfonts}
\usepackage{graphicx}
\usepackage{fancyhdr}

\usepackage[top=0.75in, bottom=0.75in, left=0.75in, right=0.75in]{geometry}
\fancypagestyle{firststyle}
{
   \fancyhf{}
   \fancyhead[L]{\textbf{\emph{To be published in The Journal of the Royal Society Interface (2014)}}}
}

\begin{document}

\title{\textbf{Mapping the stereotyped behaviour of freely-moving fruit flies}}

\author{Gordon J. Berman}
\affiliation{Joseph Henry Laboratories of Physics and Lewis-Sigler Institute for Integrative Genomics}
\author{Daniel M. Choi}
\affiliation{Department of Molecular Biology\\
Princeton University\\Princeton, NJ 08544}
\author{William Bialek}
\affiliation{Joseph Henry Laboratories of Physics and Lewis-Sigler Institute for Integrative Genomics}
\author{Joshua W. Shaevitz}\email{Corresponding author: shaevitz@princeton.edu}
\affiliation{Joseph Henry Laboratories of Physics and Lewis-Sigler Institute for Integrative Genomics}
\date{\today} 

\begin{abstract}

A frequent assumption in behavioural science is that most of an animal's activities can be described in terms of a small set of stereotyped motifs.  Here we introduce a method for mapping an animal's actions, relying only upon the underlying structure of postural movement data to organise and classify behaviours.  Applying this method to the ground-based behaviour of the fruit fly, \emph{Drosophila melanogaster}, we find that flies perform stereotyped actions roughly 50\% of the time, discovering over 100 distinguishable, stereotyped behavioural states.  These include multiple modes of locomotion and grooming.  We use the resulting measurements as the basis for identifying subtle sex-specific behavioural differences and revealing the low-dimensional nature of animal motions.

\end{abstract}

\maketitle
\thispagestyle{firststyle}
\section{Introduction}

The concept of stereotypy--that an organism's behaviours can be decomposed into discrete, reproducible elements--has influenced the study of ethology, behavioural genetics, and neuroscience for decades \cite{ALTMANN:1974uj,lehner96}.  Animals possess the ability to move in a vast continuum of ways, theoretically constrained only by the biomechanical limits of their own morphology.  Despite this, the range of behavioural actions typically performed by an animal is thought to be much smaller, constructed largely of stereotyped actions that are consistent across time, individuals, and, in some cases, even species \cite{gould82,Stephens:2011eg}.  A discrete behavioural repertoire can potentially arise via a number of mechanisms, including mechanical limits of gait control, habit formation, and selective pressure to generate robust or optimal actions.  In many instances, the search for an individual behavioural neural circuit or gene begins with the assumption that a particular action  of interest is stereotyped across time and individuals \cite{glimcher04,Manoli:2006dp}.  

Despite the centrality of this concept, with few exceptions \cite{Osborne:2005bj,Stephens:2008dk,Stephens:2011ks,Desrochers:2010he,Brown:2013ew}, the existence of stereotypy has not been probed experimentally.  This is largely due to the lack of a comprehensive and compelling mathematical framework for behavioural analysis.  Here, we introduce a new method for quantifying postural dynamics that retains an animal's full behavioural complexity, using the fruit fly \emph{Drosophila melanogaster} as a model organism to discover and map stereotyped motions.

Most prior methods for quantifying animal behaviour lie in one of two regimes.  One of these is the use of coarse metrics such as a gross activity level (e.g. mean velocity or number of times the organism crosses a barrier) or counting the relative frequencies of particular events engrained into the experimental set-up (e.g. turning left or right in a maze).  While this approach allows for high-throughput analysis of various organisms, strains, and species, only the most gross aspects of behaviour can be ascertained, potentially overlooking the often subtle effects of the manipulations of interest that are only apparent at a finer descriptive level.  The other common approach for behavioural quantification is to use a set of user-defined behavioural categories.  These categories, such as walking, grooming, or fighting,  are codified heuristically and scored either by hand or, more recently, via supervised machine-learning techniques \cite{Dankert:2009fa,Branson:2009jf,Kabra:2013jk,deChaumont:2012du,2013NatCo4E1910K}.  While the latter approach allows for higher throughput and more consistent labeling, it remains prone to human bias and anthropomorphism and often precludes objective comparisons between data sets due to the reliance on subjective definitions of behaviour.  Furthermore, these analyses assume, \textit{a priori}, that stereotyped classes of behaviour exist without first showing, from the data, that an organism's actions can be meaningfully categorised in a discrete manner. 

Ideally, a behavioural description should manifest itself directly from the data, based upon clearly-stated assumptions, each with testable consequences.  The basis of our approach is to view behaviour as a trajectory through a high-dimensional space of postural dynamics.  In this space, discrete behaviours correspond to epochs in which the trajectory exhibits pauses, corresponding to a temporally-extended bout of a particular set of motions.  Epochs that pause near particular, repeatable positions represent stereotyped behaviours.  Moreover, moments in time in which the trajectory is not stationary, but instead moves rapidly, correspond to non-stereotyped actions.  

In this paper, we construct a behavioural space for freely-moving fruit flies.  We observe that the flies exhibit approximately 100 stereotyped behaviours that are interspersed with frequent bouts of non-stereotyped behaviours.  These stereotyped behaviours manifest themselves as distinguishable peaks in the behavioural space and correspond to recognizably distinct behaviours such as walking, running, head grooming, wing grooming, etc.  Using this framework, we begin to address biological questions about the underlying postural dynamics that generate behaviour, opening the door for a wide range of other inquiries into the dynamics, neurobiology, and evolution of behaviour.

\section{Experiments}
\begin{figure}
\centering
\includegraphics[width=\columnwidth]{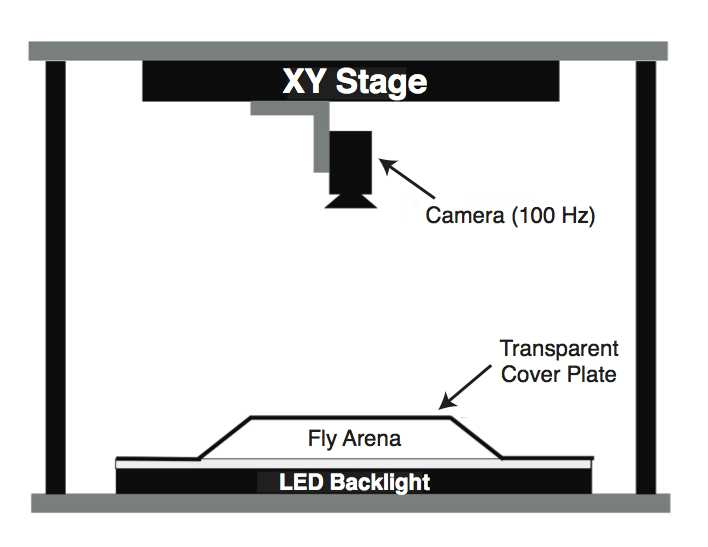}
\caption{Schematic of the imaging apparatus.}\label{apparatus}
\end{figure}

We probed the spontaneous behaviours of ground-based flies (\emph{Drosophila melanogaster}) in a largely featureless circular arena (Fig \ref{apparatus}). Under these conditions, flies display a multitude of complex, non-aerial behaviours such as locomotion and grooming, typically involving multiple parts of their bodies.  To capture dynamic rearrangements of the fly's posture, we recorded video of individual behaving animals with sufficient spatiotemporal resolution to resolve moving body parts such as the legs, wings, and proboscis. 

We designed our arena based on previous work which showed that a thin chamber with gently sloping sides prevents flies from flying, jumping, and climbing the walls \cite{Simon:2010ku}.  To keep the flies in the focal plane of our camera, we inverted the previous design. Our arena consists of a custom-made vacuum-formed, clear PETG plastic dome 100mm in diameter and 2mm in height with sloping sides at the edge clamped to a flat glass plate.  The edges of the plastic cover are sloped to prevent the flies from being occluded and to limit their ability to climb upside-down on the cover.  The underside of the dome is coated with a repellent silane compound (heptane and 1,7-dichloro-1,1,3,3,5,5,7,7-octamethylte-trasiloxane) to prevent the flies from adhering to the surface.  In practice, we find that this set-up results in no bouts of upside-down walking.  

Over the course of these experiments, we studied the behaviour of 59 male and 51 female \emph{D. melanogaster} (Oregon-R strain).  Each animal was imaged using a high-speed camera ($100\,{\rm Hz}$, $1088 \times 1088$ pixels).  A proportional-integral-derivative (PID) feedback algorithm is used to keep the moving fly inside the camera frame by controlling the position of the X-Y stage based on the camera image in real time.  In each frame we focus our analysis on a $200\times 200$ pixel square containing the fly.  We imaged each of the flies for one hour, yielding $3.6\times 10^5$ movie frames per individual, or approximately $4\times 10^7$ frames in total.  All aspects of the instrumentation are controlled by a single computer using a custom-written LabView graphical user interface.

Each of these flies was isolated within 4 hours of eclosion and imaging occurred 1-14 days after that.  Flies were placed into the arena via aspiration and were subsequently allowed 5 minutes for adaptation before data collection (Fig \ref{S1}).  All recording occurred between the hours of 9:00 AM and 1:00 PM, thus reducing the effect of circadian rhythms, and the temperature during all recordings was $25^o\pm 1 ^o C$.

\section{Behavioural analysis}
\begin{figure*}
\centering
\includegraphics[width=2\columnwidth]{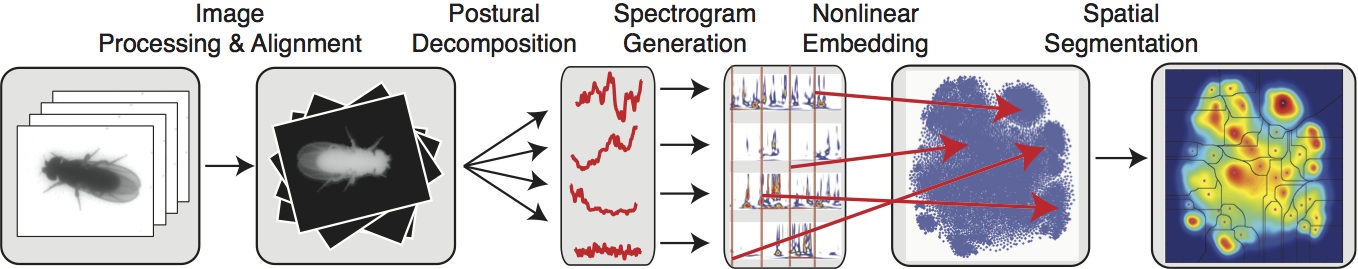}
\caption{Overview of the data analysis pipeline.  Raw images of the \emph{Drosophila melanogaster} are segmented from the background, rescaled to a reference size, and then aligned, creating a stack of images in the co-moving and co-rotating frame of the fly.  These images are then decomposed via PCA into a relatively low-dimensional set of time series.  A Morlet wavelet transform is subsequently applied to these time series, creating a spectrogram for each postural mode separately.  After normalization, each point in time is mapped into a 2-dimensional plane via t-SNE \cite{vanderMaaten:2008tm}.  Lastly, a watershed transform is applied to a gaussian-smoothed density over these points, isolating individual peaks from one another.}\label{pipeline}
\end{figure*}

The general framework of our analysis is described in Fig \ref{pipeline}.  Images are first segmented and registered in order to isolate the fly from the background and enforce translational and rotational invariance.  After this, they are decomposed into postural time series and converted into wavelet spectrograms, thus creating a spatio-temporal representation for the fly's dynamics within the images.  These spectrograms are used to construct spectral feature vectors that we embed into two dimensions using t-Distributed Stochastic Neighbor Embedding \cite{vanderMaaten:2008tm}.  Lastly, we estimate the probability distribution over this two dimensional space and identify resolvable peaks in the distribution.  We confirm that sustained pauses near these peaks correspond to discrete behavioural states.

\subsection{Image segmentation and registration}
Given a sequence of images, we wish to build a spatio-temporal representation for the fly's postural dynamics.  We start by isolating the fly within each frame, followed by rotational and translational registration to produce a sequence of images in the coordinate frame of the insect.  Details of these procedures are listed in Appendix \ref{a:images}.  In brief, we apply Canny's method for edge detection \cite{canny86}, morphological dilation, and erosion to create a binary mask for the fly.  After applying this mask, we rotationally align the images via polar cross-correlation with a template image, similar to previously developed methods \cite{decastro87,reddy96,wilson06}.  We then we use a sub-pixel cross-correlation to translationally align the images \cite{guizar08}.  Lastly, every image is re-sized so that, on average, each fly's body covers the same number of pixels.  An example segmentation and alignment is shown in Supplementary Movie S1.

\subsection{Postural decomposition}
As the fly body is made up of relatively inflexible segments connected by mobile joints, the number of postural degrees of freedom is relatively small when compared to the 40,000 pixels in each image.  Accordingly, a natural representation for the fly's posture would be to enumerate the relative angles of each of the fly's appendages as a function of time \cite{Revzen:2012cj,Ristroph:2009eb,Fontaine:2009em}.  Extracting these variables directly from the images, however, is prohibitively difficult due to occlusions and the complex fly limb and wing geometry.  

\begin{figure*}
\centering
\includegraphics[width=2\columnwidth]{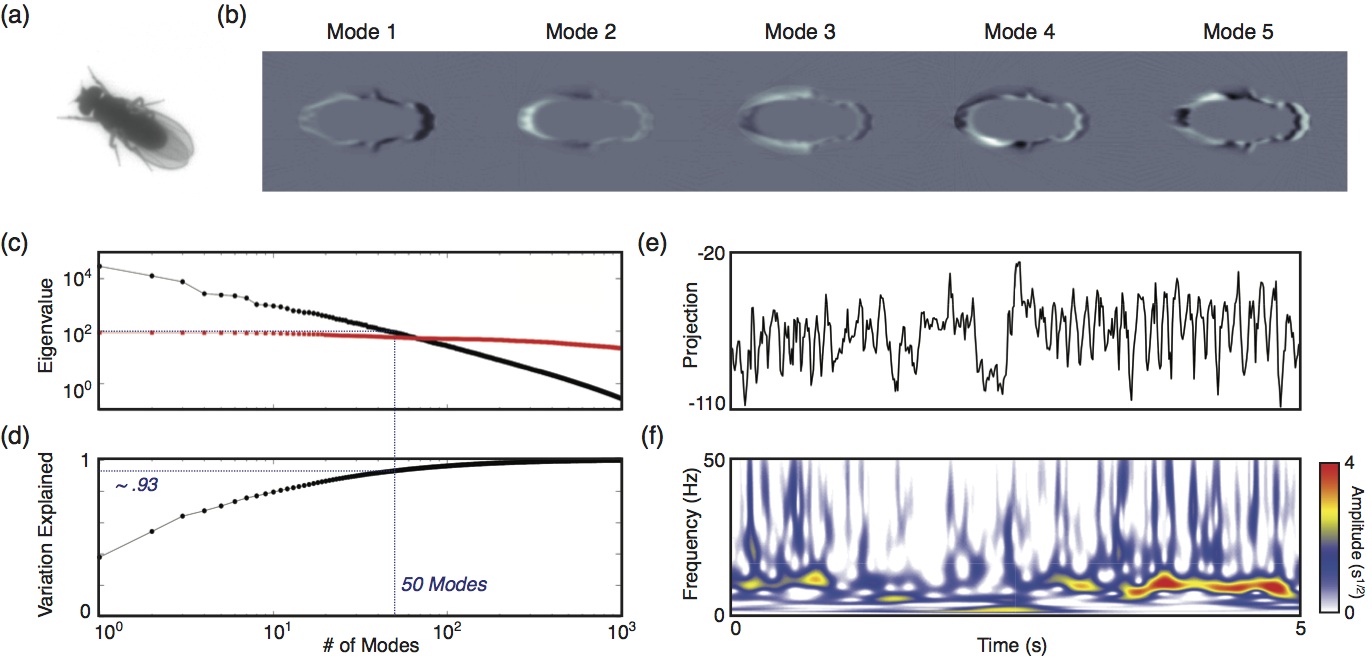}
\caption{Generation of spectral feature vectors. (a) Raw image of a fly in the arena. (b) Pictorial representation of the first 5 postural modes, $\hat{x}_{1-5}$, after inverse Radon transform.  Black and white regions represent highlighted areas of each mode (with opposite sign).  (c) First 1,000 eigenvalues of the data matrix (black) and shuffled data (red).  (d) Fraction of cumulative variation explained as a function of number of modes included.  (e) Typical time series of the projection along postural mode 6 and (f) its corresponding wavelet transform.}\label{fig1_v2}
\end{figure*}

As an alternative strategy, we find that nearly all of the variance in the $4\times 10^4$ pixel images can be explained by projecting the observed pixel values onto a Euclidean space of just 50 dimensions.  We apply Principal Component Analysis (PCA) to radon transforms of the images.  PCA is a frequently-used method for converting a set of correlated variables into a set of values of linearly uncorrelated eigenmodes. Results from this analysis can be described as the space spanned by the eigenvectors of the data covariance matrix, $C$, corresponding to the largest $m$ eigenvalues out of the total latent dimensionality of the data.  While, in general, there is no rigorous manner to choose $m$, here, we will keep all modes containing correlations larger than the finite sampling error within our data set.  According to this heuristic, we set $m = 50$ (Fig \ref{fig1_v2}(c)), a number of modes explaining approximately 93\% of the observed variation (Fig \ref{fig1_v2}(d)).  Details of this computation can be found in Appendix \ref{a:pca}.  

We refer to these directions of correlated variation as postural modes.  As seen in Fig \ref{fig1_v2}(b), these modes are fly-like in appearance, but do not lend themselves to intuitive interpretation.  However, projecting individual images onto these axes, we can convert a movie of fly behaviour into a 50-dimensional time series, 
\begin{equation}\label{posturalSpace}
\mathbf{Y} \equiv \{y_1 (t), y_2 (t), \cdots , y_{50}(t)\},
\end{equation}
as exemplified in Fig \ref{fig1_v2}(e).  

\subsection{Spectrogram generation}
The instantaneous values of the postural modes do not provide a complete description of behaviour, as our definition of stereotypy is inherently dynamical.  Previously published studies of behaviour have searched for motifs -- repeated subsequences of finite length -- within a behavioural time series \cite{Ye:2011wn,Brown:2013ew}.  However, this paradigm is often confounded by problems of temporal alignment and relative phasing between the component time series.  Additionally, certain behaviours (for example, wing grooming in \emph{Drosophila}) involve multiple appendages moving at different time scales, thus complicating the choice of motif length.  

As an alternative to this approach, we use a spectrogram representation for the postural dynamics, measuring the power at frequency $f$ associated with each postural mode, $y_k(t)$, in a window surrounding a moment in time, $S({\rm k}, f ;t)$.  More specifically, we compute the amplitudes of the Morlet continuous wavelet transform for each postural mode  \cite{morlet84}.  Although similar to a Fourier spectrogram, wavelets possess a multi-resolution time-frequency trade-off, allowing for a more complete description of postural dynamics occurring at several time scales \cite{daubechies92}.  In particular, the Morlet wavelet is adept at isolating chirps of periodic motion, similar to what we observe in our data set.  By measuring only the amplitudes of the transform, we eliminate the need for precise temporal alignment that is required in any motif-based analysis.  Details of these calculations are shown in Appendix \ref{a:morlet}, and an example spectrogram is displayed in Fig \ref{fig1_v2}(f).  For the results presented here, we look at 25 frequency channels, dyadically spaced between 1 Hz and 50 Hz, the larger of which being the Nyquist frequency of the system.

\subsection{Spatial embedding}
$S({\rm k}, f ;t)$ is comprised of 25 frequency channels for each of the 50 eigenmodes, making each point in time represented by a 1,250-dimensional feature vector encoding the postural dynamics.  As correlations, often strong, exist between the various mode-frequency channels, we expect that the dimensionality of the manifold containing the observed values of $S({\rm k}, f ;t)$ should be vastly smaller.  As such, we would like to find a low-dimensional representation that captures the important features of the data set.  

Our strategy for dimensional reduction of the feature vectors is to construct a space, $B$, such that trajectories within it pause near a repeatable position whenever a particular stereotyped behaviour is observed.  This means that our embedding should minimise any local distortions.  However, we do not require preservation of structure on longer length scales.  Hence, we chose an embedding that reduces dimensionality by altering the distances between more distant points on the manifold.  

Most common dimensionality reduction methods, including PCA, Multi-dimensional scaling, and Isomap, do precisely the opposite, sacrificing local verisimilitude in service of larger-scale accuracy \cite{cox00,Tenenbaum:2000jp,Roweis:2000ey}.  One method that does possess this property is t-Distributed Stochastic Neighbor Embedding (t-SNE) \cite{vanderMaaten:2008tm}.  Like other embedding algorithms, t-SNE aims to take data from a high-dimensional space and embed it into a space of much smaller dimensionality, preserving some set of invariants as best as possible.  For t-SNE, the conserved invariants are related to the Markov transition probabilities if a random walk is performed on the data set.  Specifically, we define the transition probability from time point $t_i$ to time point $t_j$, $p_{j\vert i}$, to be proportional to a Gaussian kernel of the distance (as of yet, undefined) between them:
\begin{equation}\label{trans_probs}
p_{j\vert i} = \frac{\exp \Bigr(-d(t_i,t_j)^2 / 2\sigma_i^2 \Bigr)}{\sum_{k\neq i} \exp \Bigr(-d(t_i,t_k)^2 / 2\sigma_i^2 \Bigr)}.
\end{equation}
All self-transitions (i.e. $p_{i\vert i}$) are assumed to be zero.  Each of the $\sigma_i$ are set such that all points have the same transition entropy, $H_i = \sum_j p_{j\vert i}\log p_{j\vert i} = 5$. This can be interpreted as restricting transitions to roughly 32 neighbors.

The t-SNE algorithm then embeds the data points in the smaller space while keeping the new set of transition probabilities, $q_{j\vert i}$, as similar to the $p_{j\vert i}$ as possible. The $q_{j\vert i}$ are defined similarly to the larger-space transition probabilities, but are now, for technical reasons, proportional to a Cauchy (or Student-t) kernel of the points' Euclidean distances in the embedded space.  This algorithm results in an embedding that minimises local distortions \cite{vanderMaaten:2008tm}.   If $p_{j\vert i}$ is initially very small or zero, it will place little to no constraint on the relative positions of the two points, but if the original transition probability is large, it will factor significantly into the cost function.

This method's primary drawback, however, is its poor memory complexity scaling ($\propto N^2$).  To incorporate our entire data set into the embedding, we use an importance sampling technique to select a training set of 35,000 data points, build the space from these data, and then re-embed the remaining points into the space as best as possible (see Appendix \ref{a:tsne} for implementation details).  

Lastly, we need to define a distance function, $d(t_i,t_j)$, between the feature vectors.  We desire this function to accurately measure how different the shapes of two mode-frequency spectra are, ignoring the overall multiplicative scaling that occurs at the beginning and the end of behavioural bouts due to the finite nature of the wavelet transform.  Simply measuring the Euclidean norm between two spectra will be greatly affected by such amplitude modulations.  However, because $S({\rm k}, f ;t)$ is composed of a set of wavelet amplitudes, it must therefore be positive semi-definite.  As such, if we define
\begin{equation}
\hat{S}(k,f;t)\equiv \frac{S(k,f;t)}{\sum_{k',f'}S(k',f';t)},
\end{equation}
then we can treat this normalised feature vector as a probability distribution over all mode-frequency channels at a given point in time.  Hence, a reasonable distance function is the Kullback-Leibler (KL) divergence \cite{coverthomas} between two feature vectors: 
\begin{eqnarray}
d(t_1,t_2) &=& D_{KL}(t_1\vert\vert t_2) \nonumber \\
&\equiv& \sum_{f,k} \hat{S}(k,f;t_1)\ \log_2 \left[\frac{\hat{S}(k,f;t_1)}{\hat{S}(k,f;t_2)}\right].
\end{eqnarray}

\section{Results}

\subsection{Embedded space dynamics}
\begin{figure*}
\centering
\includegraphics[width=2\columnwidth]{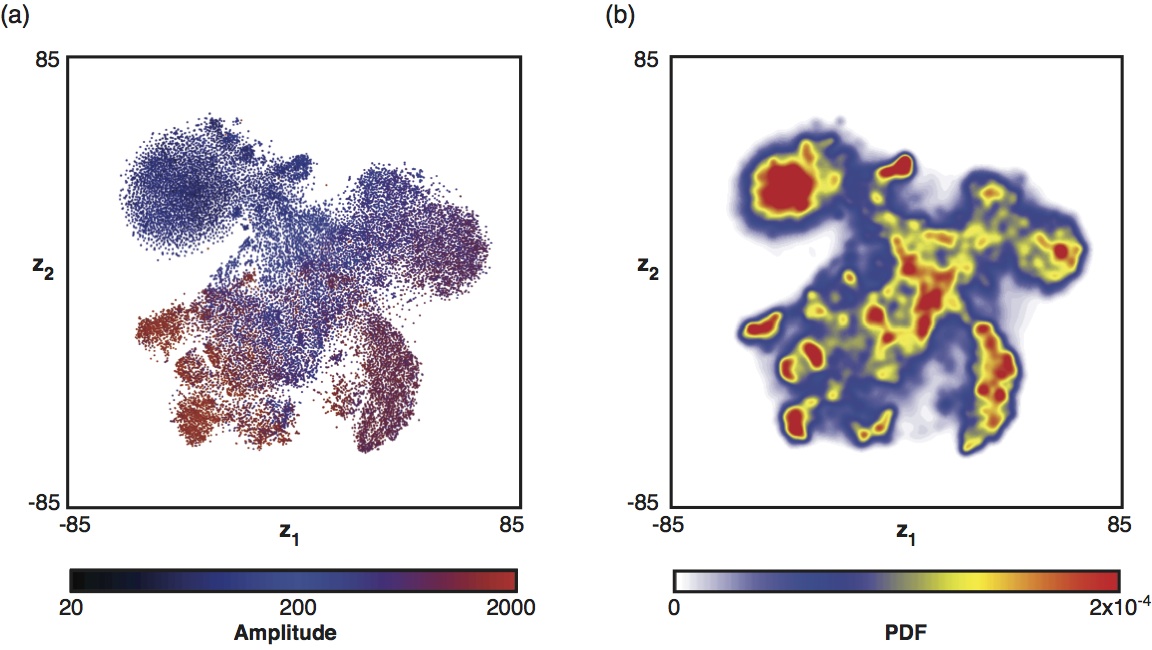}
\caption{Embedding of feature vectors.  (a) Training set points embedded into two dimensions via t-SNE.  Color coding is proportional to the logarithm of the normalization factor $\sum_{k,f}  S(k,f;t)$.  (b)  Probability density function generated from embedding all data points and convolving with a gaussian ($\sigma=1.5$). }\label{points_density}
\end{figure*}
Figure \ref{points_density} shows the embedding of our spectral feature vectors into two dimensions, the space $(z_1,z_2)$, for all of the 59 individual male flies.  We first note that nearby points have similar power ($\sum_{k,f} S({\rm k}, f ;t)$), even though the embedding algorithm normalises-out variations in the total power of the postural motions.  Embedding the same data into three dimensions yields a very similar structure with less than 2\% reduction of the embedding cost function (Eq. \ref{eq:embedcost}, Fig \ref{S3}). 

We generated an estimate of the probability density, $b(\mathbf{z})$ by convolving each point in the embedded map with a gaussian of relatively small width ($\sigma=1.5$, Fig \ref{points_density}(b)).  Far from being uniformly distributed across this space, $b(\mathbf{z})$ contains a large number of resolved local maxima.  The locations of these peaks provide a potential representation for the stereotyped behaviours that the flies perform. As expected, we find that individuals display significantly less intra- than inter-individual variation when their behavioural maps are compared (Fig \ref{S4}). 

\begin{figure}
\centering
\includegraphics[width=\columnwidth]{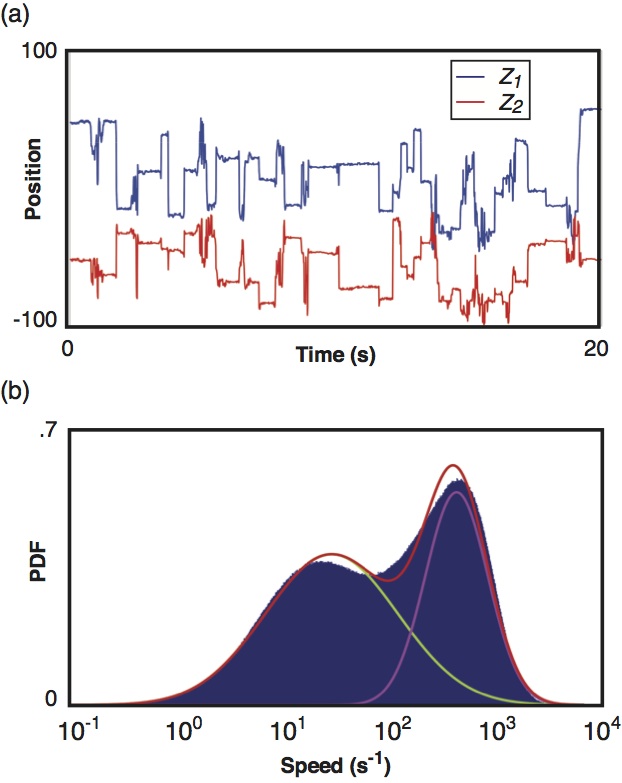}
\caption{Dynamics within behavioural space.  (a) Typical trajectory segment through behavioural space, $z_1(t)$ (blue) and $z_2(t)$ (red). (b) Histogram of velocities in the embedded space fit to a two-component log-gaussian mixture model. The blue bar chart represents the measured probability distribution, the red line is the fitted model, and the cyan and green lines are the mixture components of the fitted model.}\label{vHist}
\end{figure}

\begin{figure*}
\centering
\includegraphics[width=2\columnwidth]{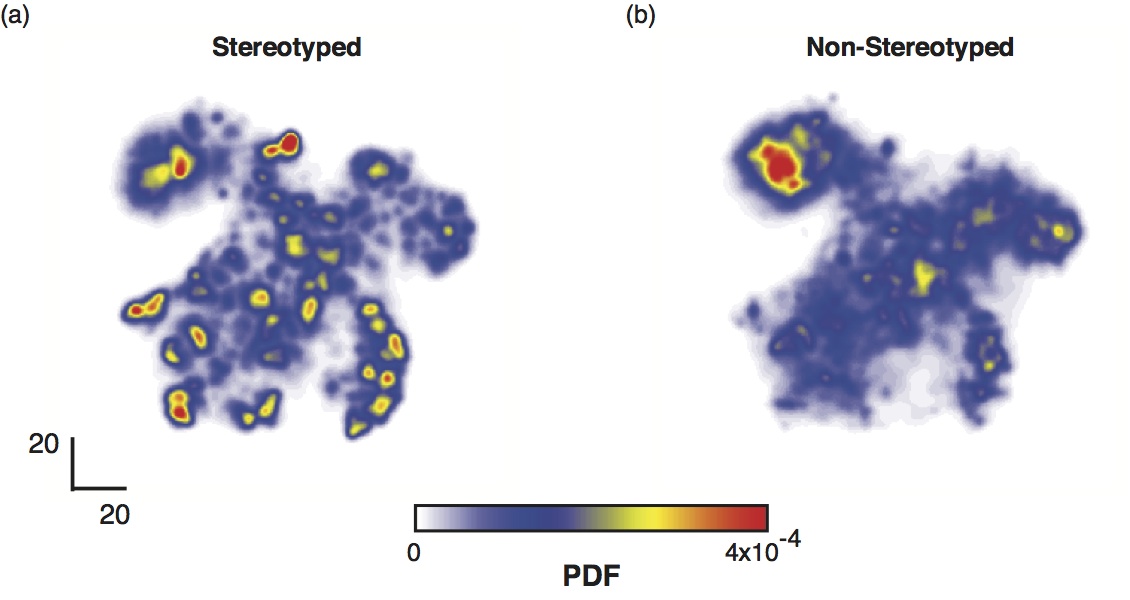}
\caption{Concentration of behavioural space during stereotyped movements.  Comparison between the densities generated during stereotyped (a) and non-stereotyped (b) epochs.}\label{stereo_non_stereo}
\end{figure*}

This space not only contains peaks, but the trajectory through it also pauses at repeatable locations.  Through numerical differentiation of $z_1(t)$ and $z_2(t)$, we observe a two-state ``pause-move'' pattern of dynamics.  Typical time traces of $z_1(t)$ and $z_2(t)$ show this type of trajectory, with long stationary periods interspersed by quick bouts of movement (Fig \ref{vHist}(a)).  More quantitatively, we find that the distribution of velocities within the embedded space is well-represented by a two-component log-normal mixture model in which the the two peaks are separated by almost two orders of magnitude (Fig \ref{vHist}(b)).  The distribution of points in the low-velocity case (approximately 45\% of all time points) is highly localized with distinguishable peaks (Fig \ref{stereo_non_stereo}).  The high-velocity points, in contrast, are more uniformly distributed.  

\subsection{Behavioural states}

\begin{figure*}
\centering
\includegraphics[width=2\columnwidth]{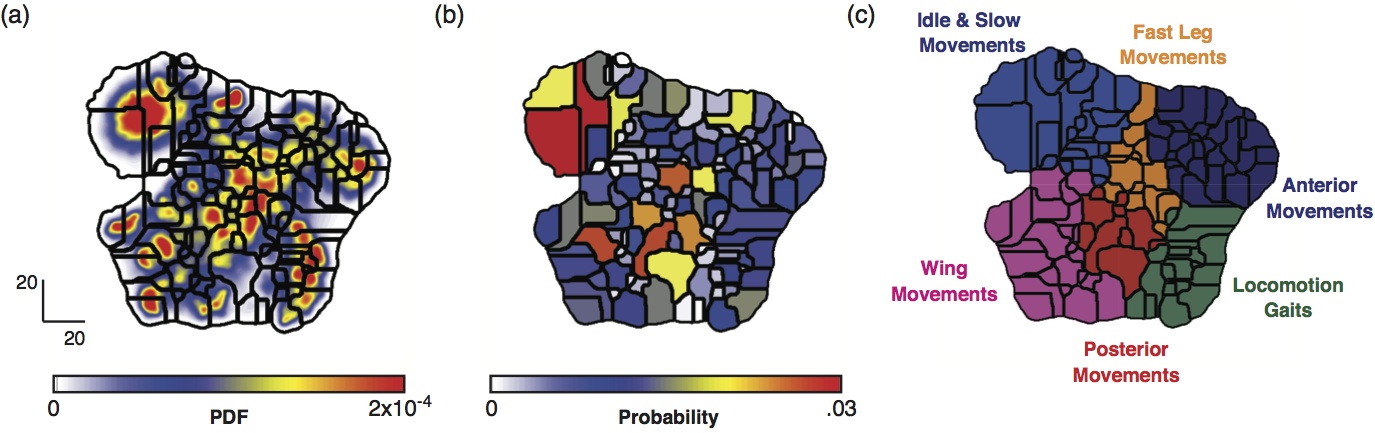}
\caption{Segmentation into behavioural regions.  (a) Boundary lines obtained from performing a watershed transform on the PDF from Fig \ref{points_density}(b).  (b) Integrated probabilities within each of the regions.  (c) The organisation of behavioural space into regions of similar movement types.  Definition of regions is performed through visual assessment of movies.}\label{region_plots}
\end{figure*}

\begin{figure}
\centering
\includegraphics[width=\columnwidth]{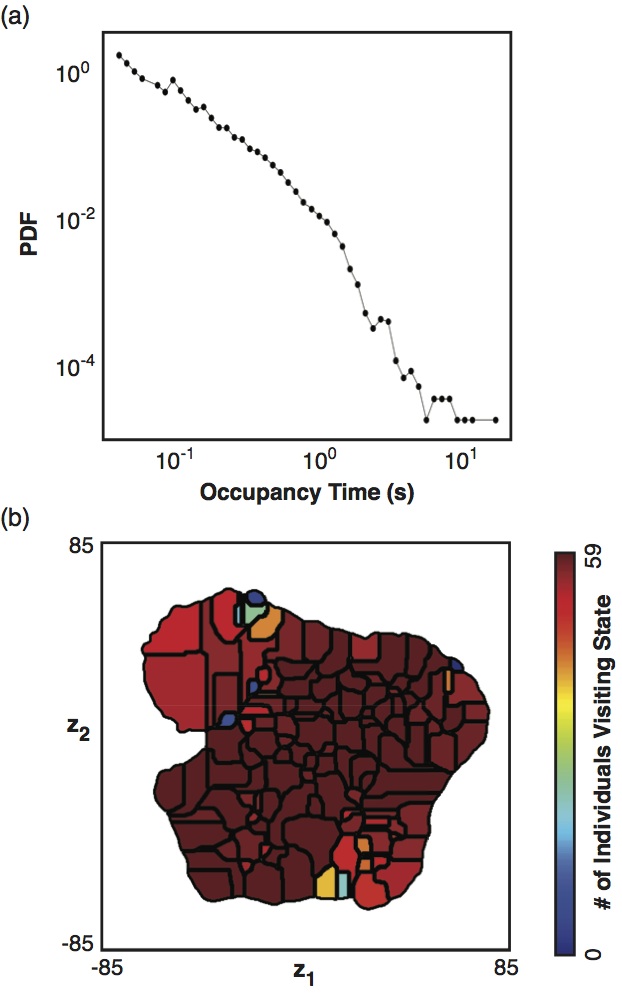}
\caption{Behavioural state dynamics.  (a)  Distribution of occupancy times in all behaviours. (b) Number of individuals (out of 59 possible) that visit each behaviour at some point during observation.}\label{visits_residents}
\end{figure}

The embedded space is comprised of peaks surrounded by valleys.  Finding connected areas in the $z_1, z_2$ plane such that climbing up the gradient of probability density always leads to the same local maximum, often referred to as a watershed transform \cite{meyer94}, we delineate 122 regions of the embedded space.  Each of these contains a single local maximum of probability density (Fig \ref{region_plots}(a)).  When the trajectory, $\mathbf{z}(t)$ pauses at one of these peaks, we find that each of these epochs correspond to the fly performing a particular stereotyped behaviour.  These pauses last anywhere from .05 s up to nearly 25 s (Fig \ref{visits_residents}(a)).

Observing segments of the original movies corresponding to pauses in one of the regions, we consistently observe the flies performing a distinct action that corresponds to a recognizable behaviour when viewed by eye (Supplementary Movies S2-11).  Many of the movements we detect are similar to familiar, intuitively defined behavioural classifications such as walking, running, front leg grooming, and proboscis extension, but here, the segmentation of the movies into behavioural categories has emerged from the data itself, not through \emph{a priori} definitions.  Moreover, we see that near-by regions of our behavioural space correspond to similar, yet distinct, behaviours (Fig \ref{region_plots}(c)).

\begin{figure}
\centering
\includegraphics[width=\columnwidth]{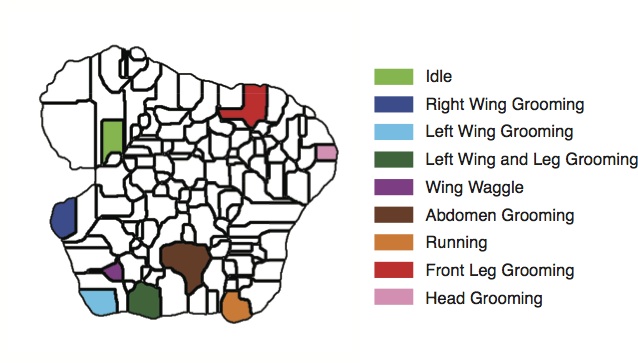}
\caption{ Behavioural space peaks correspond to specific stereotyped behaviours.  Selected regions within behavioural space are shown and are labeled via the colour-coded legend on the right.  Instances of dwells within each of these regions can be seen in Supplementary Movies 3-11.  The examples displayed in these movies are randomly selected and contain clips from many different flies, showing that the behavioural space provides a representation of behaviour that is consistent across individuals.}\label{movie_region}
\end{figure}

This classification is consistent across individuals (Figures \ref{visits_residents}-\ref{movie_region}, Supplementary Movies S3-11).  The vast majority of these regions are visited by almost all of the flies at some point (Fig \ref{visits_residents}(b)).  104 of the 122 regions were visited by over 50 (of 59 total) flies, and the remaining behaviours were all low-probability events, containing, in total, less than 3\% of the overall activity.  

\subsection{Behavioural states as periodic orbits}
\begin{figure*}
\centering
\includegraphics[width=2\columnwidth]{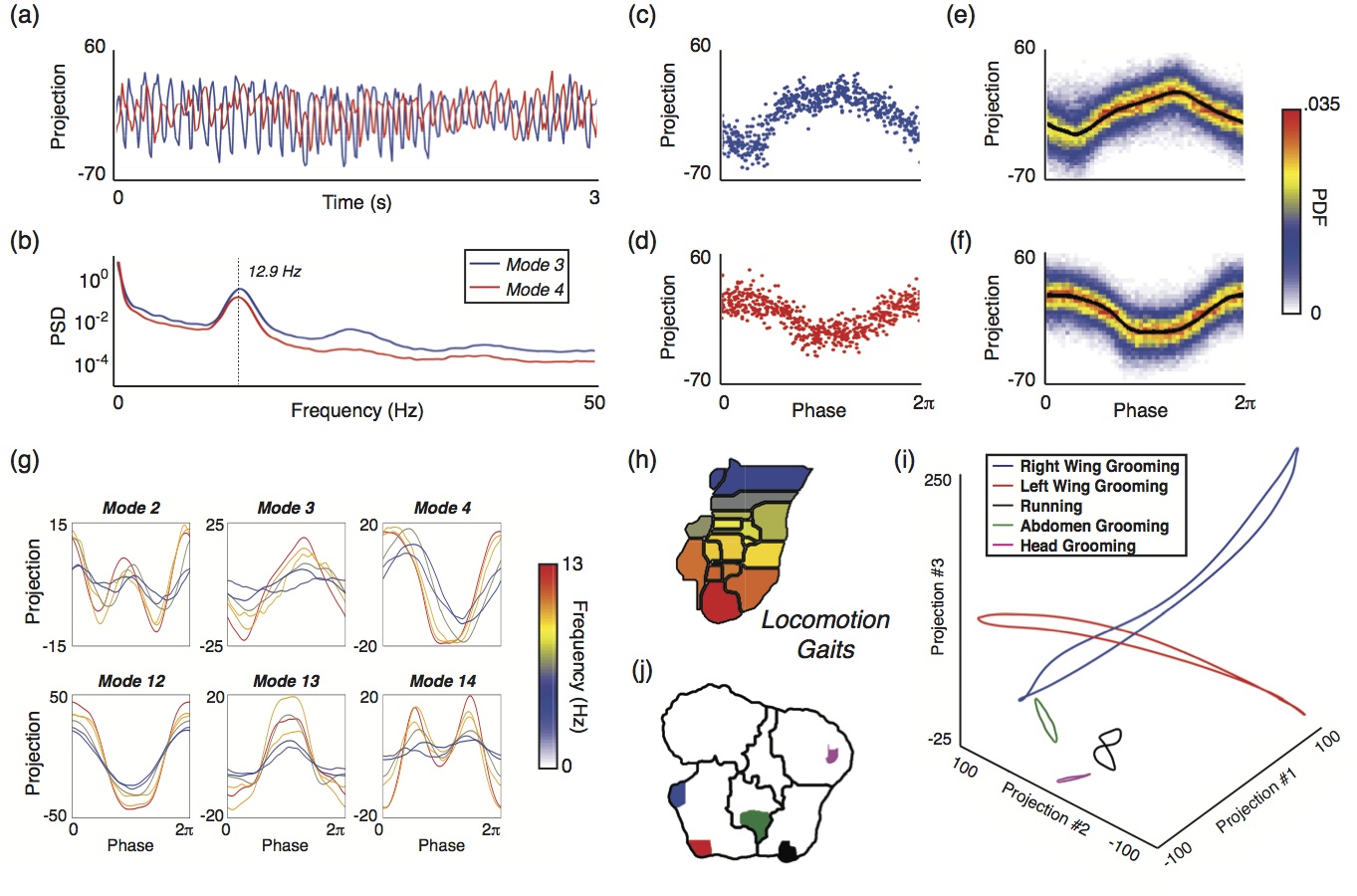}
\caption{Periodic dynamics within behavioural states.  (a) Periodic oscillations in the 3rd and 4th postural eigenmodes during a typical high-frequency running sequence. (b) Average power spectral density for all instances of this behaviour (the bottom-most region in (h)). (c)-(d) Phase reconstruction of the data in (a) for modes three and four (respectively) .  (e)-(f) Probability densities of projections along the 3rd and 4th modes (respectively) for all instances of the behaviour shown in (a)-(d).  The black line is the phase-averaged curve (via Eqn \ref{phase_average}).  (g) Comparison between the phase-averaged curves for 7 different locomotion gaits.  Line colours are proportional to the mean gait frequency.  (h) Locomotion gaits from Fig \ref{region_plots}(c), colour-coded by mean frequency.  The colour scale here is the same as in (g).  (i) 3-D plots of the phase-averaged trajectories for 5 different behaviours.  The first three postural modes are plotted here.  (j) Regions corresponding to the orbits shown in (i) (coded by color).}\label{orbit_figure}
\end{figure*}

Periodic orbits in postural movements are suggestive of underlying low-dimensional dynamic attractors that produce stable behavioural templates \cite{Full:1999vc}. These types of motifs have been hypothesized to form the basis for neural and mechanical control of legged locomotion at fast time scales \cite{Holmes:2006uf}.  Because our behavioural mapping algorithm is based upon similarities between postural frequencies exhibited at different times, a potential hypothesis is that pauses in behavioural space correspond to periodic trajectories in the space of postural movements (Eqn \ref{posturalSpace}). In our data, a fast running gait (the bottom-most region of Fig \ref{orbit_figure}(h)) corresponds to periodic oscillations of the postural time series with a clear peak at 12.9 Hz in the power spectral density (Fig \ref{orbit_figure}(a)-(b)). This frequency is in good agreement with previous measurements of the fly walking gait \cite{Strauss:1990ua,Wosnitza:2013il}.

To systematically investigate the periodicity of the postural dynamics, for each behavioural bout we map time onto a phase variable, a cyclic coordinate defined on the unit circle.  This process is usually referred to as phase reconstruction.  The method we use, \emph{Phaser} \cite{Revzen:2008hb}, performs Hilbert transforms to construct phase estimations from several time series separately, then combines these estimates via a maximum likelihood estimate that uses Fourier-series based corrections.  Here, we apply \emph{Phaser} to the postural mode time series, $y_k(t)$, treating the correlated motions along all 50 postural eigenmodes as synchronized oscillators.  We performed this reconstruction for each multi-cycle behavioural bout.  After reconstructing the phases for all of the 5,483 bouts of fast running observed in male flies, we observe a clear periodic pattern across several of the postural modes (Fig \ref{orbit_figure}(c)-(f)).

This type of analysis also brings additional insight into the subtle distinctions between our observed behavioural states.  If we construct phase-averaged orbits for seven of the running behaviours, we observe many differences in the gait dynamics (see Appendix \ref{a:phases}, Fig \ref{orbit_figure}(g)).  For instance, we observe an increase in many mode amplitudes as the gait frequency increases (e.g. in modes 3, 12, and 13), as noted in previous work \cite{Mendes:2013cu}.  In addition, we also see subtle changes in phase (e.g. in mode 4), as well as a near-elimination of a period-doubled trajectory (seen in mode 14).  This type of observation could allow for a more thorough understanding of speed control and gait transitions in hexapod locomotion.

We also find oscillatory postural dynamics for other stereotyped behaviours, with many behaviours resulting in a periodic orbit in postural space (Fig \ref{orbit_figure}(i)).  These behaviours are found in many regions of behavioural space, suggesting that much of behaviour is indeed confined to low-dimensional postural dynamics.  It is important to note that periodic trajectories emerge directly from our analysis, even though the wavelet transform used to define our feature vectors does not preserve phase information.

\subsection{Differences in behaviour between males and females}
\begin{figure*}
\centering
\includegraphics[width=2\columnwidth]{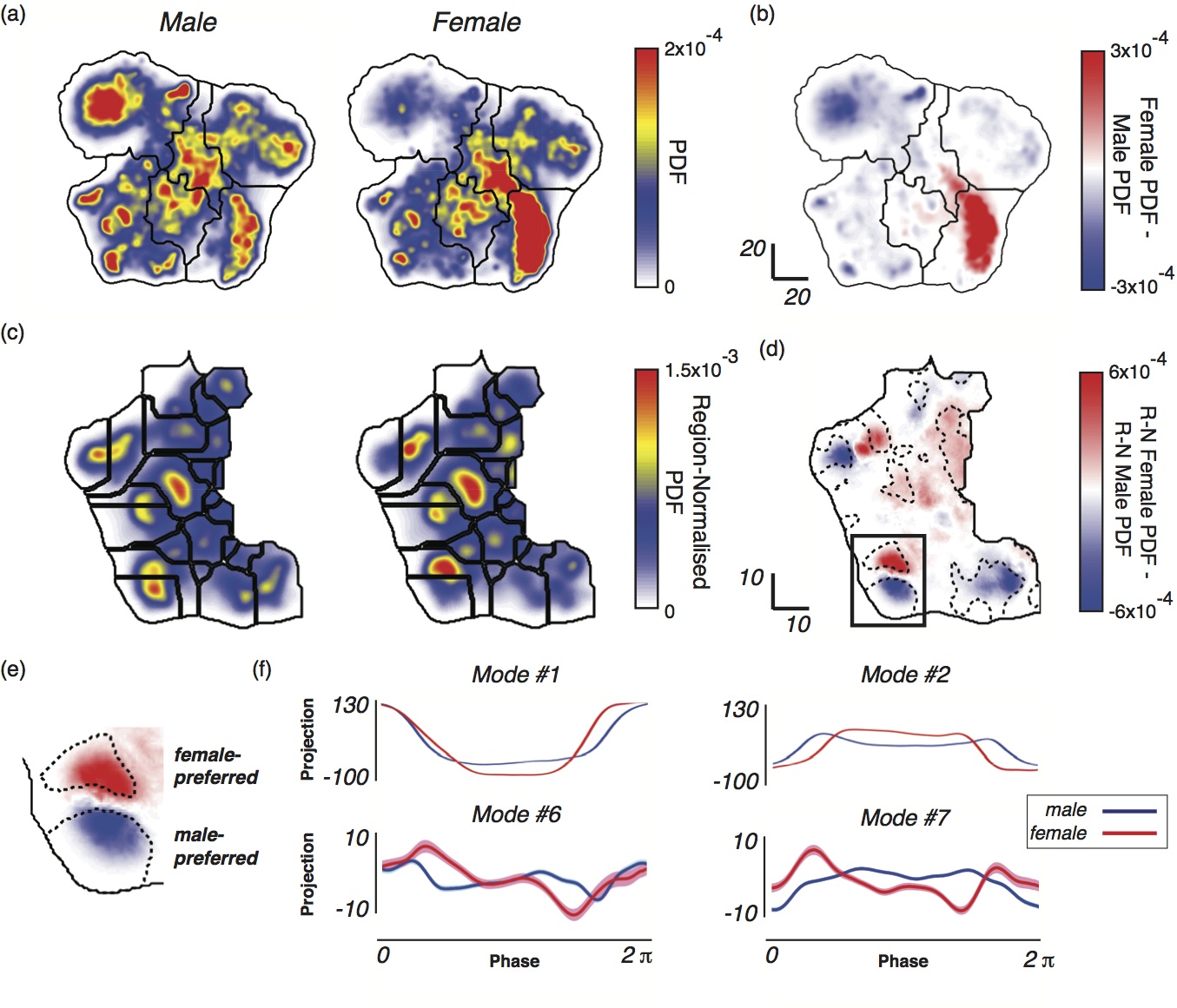}
\caption{Comparison between male and female behaviours.  (a) Measured behavioural space PDF for male (left) and female (right) flies.  (b) Difference between the two PDFs in (a).  Here we observe large dimorphisms between the sexes, particularly in the "Locomotion Gaits" and "Idle and Slow Movements" regions.  (c) PDFs for behaviours in the ``Wing Movements'' portion of the behavioural space (the lower left of the full space).  These PDFs (male on the left and female on the right) are normalised so that they each integrate to one.  The black lines are the boundaries found from a watershed transform and are included to guide the eye.  (d) Difference between the two normalised behavioural spaces in (c).  Dashed lines enclose regions in which the median male and the median female PDF values are statistically different via the Wilcoxan rank sum test ($p<.01$). (e) Zoom-in on the boxed region in (d).  Both of these regions correspond to left wing grooming, but with behaviours within the male-preferred region incorporating an additional leg kick (Supplementary Movies S12-13).  (f) Average periodic orbits for postural eigenmodes 1, 2, 6, and 7.  The area surrounding the lines represents the standard error of the mean at each point along the trajectory.  Average periodic orbits for all of the first 25 postural modes are shown in Fig \ref{S6}. }\label{male_female_figure}
\end{figure*}

To demonstrate the power of this method to detect subtle differences in behaviour, we compared the behavioural spaces of male and female fruit flies by embedding the postural time series data from females into the behavioural space derived from the male flies (Fig \ref{points_density}). Figure \ref{male_female_figure}(a) displays the male and female behavioural probability densities.  We find a striking difference between the two sexes, with locomotory behaviours  greatly enhanced but resting and slow motions largely suppressed in females when compared to males.  This is in agreement with previous results, showing that young females are more active than their male counterparts \cite{LeBourg:1987wb}.

We then sought to isolate subtle behavioural differences between the sexes that are evident in the fine-scale structure of these maps.  An example of this can be seen in the ``Wing Movements'' portion of the behavioural space (the lower left corner of the map).  First, we obtained both male and female region-normalised (R-N) PDFs (Fig, \ref{male_female_figure}(c)), where the integral of the behavioural space density within the "Wing Movements" region integrates to one. Within the space of wing movements, we identified regions that show statistically significant differences between the two sexes using a Wilcoxon rank sum test \cite{Wilcoxon45} at each point in behavioural space. This test determines the locations of significant difference between the median male PDF value and the median female PDF value ($p$-value $<.01$).  Regions where significant differences were found are indicated by the dashed lines in Figure \ref{male_female_figure}(d).

Particular behaviours, such as left-wing grooming, are sexually dimorphic (\ref{male_female_figure}(d) solid box, Movies S12-S13). Male-preferred grooming  includes a kick of the middle leg on the left side of the body that clears the profile of the wing and moves anteriorly before pushing back towards the posterior.  Female-preferred grooming lacks this additional leg movement. We verified this differences  by isolating the mean postural-space orbits associated with each of these regions (Figures \ref{male_female_figure}(f), \ref{S6}). Importantly, while these orbits are statistically different, the average frequencies for the behaviours are not ($f_{\rm male}=3.49\pm.15$ Hz versus $f_{\rm female}=3.28\pm.08$ Hz). We note that these results are consistent across a large range of the behavioural-map smoothing parameter $\sigma$ (Fig \ref{S5}), such that fine-tuning of the spatial structure of the behavioural map is not necessary to obtain the results seen here.

It should be noted that future study is necessary to determine the ethological relevance of these findings and to understand how much of the variance we observe is related to the specifics of our experimental paradigm.  However, the fact that these distinctions are found without specifically looking for any of them -- emerging only from underlying statistics of the behavioural map -- provides quantitative verification that the classifications we make are meaningful.  Inherent in any unsupervised classification method is the question of how to validate its accuracy.  Here, there is no ground truth with which to compare, since a significant aim of our work is to dispense with \emph{a priori} behavioural definitions.  However, by showing that meaningful distinctions and agglomerations can be made between different behavioural instances, we provide evidence that the approach introduced here can become the basis undergirding a wide range of experimental investigations into the behaviour of animals.

\section{Conclusions}
The ability to map and compare the behavioural repertoire of individuals and populations of animals has applications beyond the study of terrestrial dynamics in fruit flies. Combined with tools for genetic manipulation, DNA sequencing, neural imaging, and electrophysiology, the identification of subtle behavioural distinctions and patterns between groups of individuals will impact deep questions related to the interactions between genes, neurons, behaviour, and evolution. In this initial study, we probed the motion of individuals in a largely featureless environment. Extensions to more complicated situations, e.g. where sensory inputs are measured and/or controlled, genes are manipulated, or multiple individuals are present, are readily implemented.  

Finally, we note that the only \emph{Drosophila}-specific step in our analysis pipeline is the generation of the postural eigenmodes. Given movies of sufficient quality and length from different organisms, spectral feature vectors and behavioural spaces can be similarly generated, allowing for potential applications from worms to mice to humans and a greater understanding of how animals behave.

\section*{Acknowledgements}
 We thank Yi Deng, Kieren James-Lubin, Kelsi Lindblad, and Ugne Klibaite for assistance in data collection and analysis, as well as Jessica Cande, David Stern, and David Schwab for discussions and suggestions.  JWS and GJB also acknowledge the Howard Hughes Medical Institute Janelia Farm Visitor Program and the Aspen Center for Physics, where many ideas for this work were formulated. This work was funded through awards from the National Institutes of Health (GM098090, GM071508), The National Science Foundation (PHY--0957573, PHY--1066293), the Pew Charitable Trusts, the Swartz Foundation, and the Alfred P. Sloan Foundation.
\section*{Data Accessibility}
Example code for running the algorithms can be found at https://github.com/gordonberman/MotionMapper.  For access to raw image data, please email JWS at shaevitz@princeton.edu.

\appendix

\section*{Appendices}

\section{Image processing}\label{a:images}
To isolate the fly from the background, we apply Canny's method for edge detection \cite{canny86}, resulting in a binary image containing the edge positions.  We then morphologically dilate this binary image by a $3\times 3$ square in order to fill any spurious holes in the edges and proceed to fill all closed curves.  This filled image is then morphologically eroded by a square of the same size, resulting in a mask.  After applying this mask to the original image, we now have our segmented image.

While our tracking algorithm ensures that the fly remains within the image boundaries, the centre of the fly and the orientation within the frame vary over time.  Having obtained a sequence of isolated fly images, we next register them both translationally and rotationally with respect to a template image.  The template image is generated by taking a typical image of a fly and then manually ablating the wings and legs digitally.

For our first step, we rotationally align.   This is achieved through finding the angle that maximises the cross-correlation between the magnitudes of the 2D polar Fourier transforms for each image and the template.   Because all translation information appears in the phase of the 2D Fourier transform, this rotational alignment, based only upon the magnitude of the transform, is independent of any initial translations between the images.  Accordingly, once rotational alignment is achieved, we can subsequently register the images translationally via a cross-correlation.

\section{Postural decomposition from images}\label{a:pca}
The aim of the postural decomposition is to take our set of $200\times 200$ aligned images and create a lower-dimensional representation that can be made into time series.  Naively, one would simply perform PCA on the the images, using each pixel value as a separate dimension.  The fly images, however, contain too many pixels to analyse due to memory limitations.  

To make this problem more tractable, we analyse only the subset of these pixels which have non-negligible variance.  Many pixels within the fly image are either always zero or always saturated, thus containing almost no dynamical information. Accordingly, we would like to use only a subsample of these measurements.  The most obvious manner to go about this is to find the pixels containing the highest variance and keep only those above a certain threshold.  The primary difficulty here, however, is that there is not an obvious truncation point (Fig \ref{S2}A).  This is most likely the result of the fact that the fly legs can potentially occupy the majority of the pixels in the image but only are present in a relatively small number in any given frame.  Hence, many of these periphery pixels all have similarly moderate standard deviations, making them difficult to differentiate.

A more compact scheme is to represent the images in Radon-transform space, which more sparsely parameterises lines such as legs or wing veins.  After Radon transformation, the probability density function of pixel-value standard deviations has a clear minimum and we keep pixels whose standard deviation is larger than this value (Fig \ref{S2}B).  This results in keeping 6,763 pixels out of 18,090, retaining approximately 95\% of the total variation in the images.  If there are $N$ images in our sample, we can represent our data set, $\mathbf{X}$ as an $N\times 6,763$-element matrix.  We then proceed to calculate the principal directions of variation in these data using PCA, as seen in Fig \ref{fig1_v2}.

Lastly, the question remains of how many modes to keep in our analysis, a task made more ambiguous due to the smoothness of the eigenvalue spectrum.  Our approach to determining the truncation point is to compare the PCA eigenvalues with a null model based on the noise properties of our data set.  Specifically, we assume that the noise is due to finite data collection.  Although additional errors in image segmentation and registration assuredly exist in our data set, this set null model provides an upper-bound on the number of statistically meaningful eigenmodes.  

To calculate this truncation point, we take our data matrix, $\mathbf{X}$, and shuffle each of its columns independently from one another, hence eliminating all meaningful correlations between them.  Given finite sampling (even if very large), however, there will still remain some residual correlations, resulting in off-diagonal non-zero terms in the covariance matrix.  Hence, if we diagonalise this new covariance matrix, the largest eigenvalue provides a resolution limit for our ability to distinguish signal from finite sampling noise.  Performing this analysis, we find that only 50 modes have eigenvalues larger than this largest shuffled eigenvalue.  These 50 modes account for slightly more than 93\% of the observed variation in the data.

\section{Wavelet calculations}\label{a:morlet}
We use the Morlet continuous wavelet transform to provide a multiple time scale representation of our postural mode dynamics.  More explicitly, we calculate this transform, $W_{s,\tau}[y(t)]$, via

\begin{equation}\label{wavelet}
W_{s,\tau}[y(t)] = \frac{1}{\sqrt{s}}\int_{-\infty}^\infty y(t) \psi^*\Bigr(\frac{t-\tau}{s}\Bigr)dt,
\end{equation}
with 
\begin{equation}\label{morlet}
\psi(\eta) =  \pi^{-1/4}e^{i \omega_0 \eta}e^{-\frac{1}{2}\eta^2}.
\end{equation}
Here, $y_i(t)$ is a postural time series, $s$ is the time scale of interest, $\tau$ is a point in time, and $\omega_0$ is a non-dimensional parameter (set to 5 here).  

The Morlet wavelet has the additional property that the time scale, $s$, is related to the Fourier frequency, $f$,  by
\begin{equation}\label{fourierfreq}
s(f) = \frac{\omega_0 + \sqrt{2+\omega_0^2}}{4\pi f}.
\end{equation}
This can be derived by maximizing  the response to a pure sine wave, $A(s,f)\equiv\Big\vert W_{s,\tau} [e^{2\pi i f t}]\Big\vert$, with respect to $s$. 

However, $A(s,\omega)$ is disproportionally large when responding to pure sine waves of lower frequencies.  To correct for this, we find a scalar function $C(s)$ such that
\begin{equation}
C(s)A(s,\omega^*) = 1 \mbox{ for all } s, 
\end{equation} 
where $\omega^*$ is  $2\pi$ times the Fourier frequency found in  Eq. \ref{fourierfreq}.  For a Morlet wavelet, this function is
\begin{equation}
C(s) = \frac{\pi^{-\frac{1}{4}}}{\sqrt{2s}}e^{\frac{1}{4}\Big(\omega_0-\sqrt{\omega_0^2+2}\Big)^2}.
\end{equation}
Accordingly, we can define our power spectrum, $S({\rm k}, f ;t)$, via 
\begin{equation}\label{fullWavelet}
S({\rm k}, f ;\tau) = \frac{1}{C(s(f))}\Big\vert W_{s(f),\tau}[y_k(t)] \Big\vert
\end{equation}

Last, we use a dyadically-spaced set of frequencies between $f_{min} = 1$ Hz and the Nyquist frequency ($f_{max} = 50$ Hz) via
\begin{equation}
f_i = f_{max} 2^{-(i-1)/(N_f-1)\log_2 \frac{f_{max}}{f_{min}}}
\end{equation}
for $i=1,2,\ldots,N_f$ (and their corresponding scales via Eq. \ref{fourierfreq}).  This creates a wavelet spectrogram that is resolved at multiple time-scales for each of the first 50 postural modes.

\section{t-SNE implementation}\label{a:tsne}
For our initial embedding using t-SNE, we largely follow the method introduced in \cite{vanderMaaten:2008tm}, minimizing the cost function 
\begin{equation}
C = D_{KL}(P \vert\vert Q) = \sum_{ij} p_{ij} \log\frac{p_{ij}}{q_{ij}}, \label{eq:embedcost}
\end{equation}
where $p_{ij} = \frac{1}{2}(p_{j\vert i}  + p_{i\vert j})$, 
\begin{equation}
q_{ij} = \frac{(1 + \Delta_{ij}^2)^{-1}}{\sum_k\sum_{\ell \ne k} (1 + \Delta_{k,\ell}^2)^{-1}},
\end{equation}
and $\Delta_{ij}$ is the Euclidean distance between points $i$ and $j$ in the embedded space.  The cost function is optimised through a gradient descent procedure that is preceded by an early-exaggeration period, allowing for the system to more readily escape local minima.  

The memory complexity of this algorithm prevents the practical number of points from exceeding $\approx 35,000$.  Although improving this number is the subject of current research \cite{2013arXiv1301.3342V}, our solution here is to generate an embedding using a selection of roughly 600 data points from each of the 59 individuals observed (out of $\approx 360,000$ data points per individual).  To ensure that these points create a representative sample, we perform t-SNE on 20,000 randomly-selected data points from each individual.  This embedding is then used to estimate a probability density by convolving each point with a 2D gaussian whose whose width is equal to the distance from the point to its $N_{embed} =10$ nearest neighbours.  This space is segmented by applying a watershed transform \cite{meyer94} to the inverse of the PDF, creating a set of regions.  Finally, points are grouped by the region to which they belong and the number of points selected out of each region is proportional to the integral over the PDF in that region.  This is performed for all data sets, yielding a total of 35,000 data points in the training set.

Given the embedding resulting from applying t-SNE to our training set, we wish to embed additional points into our behavioural space by comparing each to the training set individually.  Mathematically, let $X$ be the set of all feature vectors in the training set, $X'$ be their associated embeddings via t-SNE, $z$ be a new feature vector that we would like to embed according to the mapping between $X$ and $X'$, and $\zeta$ be the embedding of $z$ that we would like to determine.

As with the t-SNE cost function, we will embed $z$ by enforcing that its transition probabilities in the two spaces are as similar as possible.  Like before, the transitions in the full space, $p_{j\vert z}$, are given by 
\begin{equation}
p_{j\vert z} = \frac{\exp \Bigr(-d(z,j)^2 / 2\sigma_z^2 \Bigr)}{\sum_{x\in X} \exp \Bigr(-d(z,k)^2 / 2\sigma_z^2 \Bigr)},
\end{equation}
where $d(z,j)$ is the Kullback-Leibler divergence between $z$ and $x \in X$, and $\sigma_z$ is once again found by constraining the entropy of the condition transition probability distribution, using the same parameters as for the t-SNE embedding.  Similarly, the transition probabilities in the embedded space  are given by
\begin{equation}\label{E:tdist}
q_{j\vert \zeta} = \frac{(1 + \Delta_{\zeta,j}^2)^{-1}}{\sum_{x'\in X'} (1 + \Delta_{\zeta,x'}^2)^{-1}},
\end{equation}
where $\Delta_{\zeta,x'}$ is the Euclidean distance between $\zeta$ and $y\in X'$.  

For each $z$, we then seek the $\zeta^*$  that minimises the Kullback-Leibler divergence between the transition probability distributions in the two spaces:
\begin{eqnarray}\label{KLdiv}
\zeta^* &=& \arg\min_{\zeta} D_{KL}(p_{x\vert z} \vert\vert q_{y\vert\zeta}) \\
&=& \arg\min_{\zeta} \sum_{x\in X} p_{x\vert z} \log \frac{p_{x\vert z} }{q_{y(x)\vert \zeta} }.
\end{eqnarray}
As before, this is a non-convex function, leading to potential complexities in performing our desired optimization.  However, if we start a local optimization (using the Nelder-Mead Simplex algorithm \cite{jongen04,lagarias98}) from a weighted average of points, $\zeta_0$, where
\begin{equation}\label{wavg}
\zeta_0 = \sum_{x\in X} p_{x\vert z} y(x),
\end{equation}
this point is almost always within the basin of attraction of the global minimum.  To ensure that this is true in all cases, however, we also perform the same minimisation procedure, but starting from the point $y(x^*)$, where 
\begin{equation}
x^* = \arg\max_x p_{x\vert z}.
\end{equation}
This returned a better solution approximately 5\% of the time.  

Because this embedding can be calculated independently for each value of $z$, the algorithm scales linearly with the number of points.  We also make use of the fact that this algorithm is embarrassingly parallelizable.  Moreover, because we have set our transition entropy, $H$, to be equal to 5, there are rarely more than 50 points to which a given $z$ has a non-zero transition probability.  Accordingly, we can speed up our cost function evaluation considerably by only allowing $p_{x\vert z} > 0$ for the nearest 200 points to $z$ in the original space.

Lastly, we find the space of behaviours for the female data sets by embedding these data into the space created with the male training set.  We find that the median re-embedding cost (Eqn. \ref{KLdiv}) for the female cost is only 1\% more than the median re-embedding cost for the male data (5.08 bits vs. 5.12 bits) indicating that the embedding works well for both sexes.

\section{Phase-averaged orbits}\label{a:phases}
After applying the \emph{Phaser} algorithm, we find the phase-averaged orbit via a von Mises distribution weighted average.  More precisely, we construct the average orbit for eigenmode $k$, $\mathbf{\mu} ^{(k)}(\phi)$ via 
\begin{equation}\label{phase_average}
\mathbf{\mu}^{(k)} (\phi) = \sum_i y^{(k)}_i \frac{\exp [\kappa \cos (\phi - \phi_i)]}{\sum_j\exp [\kappa \cos (\phi - \phi_j )]},
\end{equation}
where $y^{(k)}_i$ is the projection onto the $k^{th}$ eigenmode at time point $t_i$, $\phi_i$ is the phase associated with the same time point, and $\kappa$ is related to the standard deviation of the von Mises distribution ($\sigma_{vM}^2(\kappa) = 1 - \frac{I_1(\kappa)}{I_0 (\kappa)}$, where $I_\nu (x)$ is the modified Bessel function of $\nu^{th}$ order).  Here we  find the value of $\kappa\approx 50.3$, which is the $\kappa$ resulting in $\sigma_{vM} = .1$.  

Because phase reconstruction only is unique up to an additive constant, to compare phase-averaged curves of different behavioural bouts, an additional alignment needs to occur.  This is performed by first finding the maximum value of cross-correlation between the phase-averaged curves for each mode.  Then, the phase offset between that pair of 50-dimensional orbits is given by the median of these found phase shifts.

\clearpage
\onecolumngrid
\renewcommand{\theequation}{S\arabic{equation}}
\setcounter{figure}{0}\renewcommand{\thefigure}{S\arabic{figure}}
\renewcommand{\thetable}{S\arabic{table}}

\section{Supplementary Tables}

\begin{table}[ht]
\caption{Parameters used in eigen-decomposition \label{parameters:pca}}
\begin{center}
\begin{tabular}{| c | l | c |}
\hline
\textbf{Parameter} & \textbf{Description} & \textbf{Value} \\
\hline
$N_{\theta}$ & Number of angles used in Radon transforms & 90 \\ \hline 
$M$ & Number of postural eigenmodes  (found, not defined \emph{a priori})& 50 \\ \hline 
\end{tabular}
\end{center}
\end{table}

\begin{table}[ht]
\caption{Parameters used in wavelet analysis\label{parameters:wavelets}}
\begin{center}
\begin{tabular}{| c | l | c |}
\hline
\textbf{Parameter} & \textbf{Description} & \textbf{Value} \\
\hline
$N_f$ & Number of frequency channels & 25 \\ \hline 
$\omega_0$ & Non-dimensional Morlet wavelet parameter & 5 \\ \hline 
$f_{min}$ & High-pass frequency cut-off & 1 Hz \\ \hline 
$f_{max}$ & Low-pass frequency cut-off & 50 Hz \\ \hline 
\end{tabular}
\end{center}
\end{table}

\begin{table}[ht]
\caption{Parameters used in t-SNE implementation\label{parameters:tsne}}
\begin{center}
\begin{tabular}{| c | l | c |}
\hline
\textbf{Parameter} & \textbf{Description} & \textbf{Value} \\
\hline
$H$ & Transition entropy & 5 \\ \hline 
$N_{train}$ & Training set size & 35,000 \\ \hline 
$N_{neighbors}$ & Maximum non-zero re-embedding transitions & 200 \\ \hline 
$N_{embed}$ & Nearest neighbors for training set calculation & 10 \\ \hline 
\end{tabular}
\end{center}
\end{table}

\begin{table}[ht]
\caption{Parameters used in behavioural segmentation \label{parameters:cluster}}
\begin{center}
\begin{tabular}{| c | l | c |}
\hline
\textbf{Parameter} & \textbf{Description} & \textbf{Value} \\ \hline 
$\sigma$ & Width of gaussian kernel density estimator for embedded points & 1.5 \\ \hline 
$\tau_{min}$ & Minimum behavioural time scale & $.05$ s  \\ \hline
$\sigma_{v}$ & Width of gaussian for embedded-space velocity calculations & $.02$ s  \\ \hline
$\sigma_{vM}$ & Standard deviation of von Mises smoothing for phase averaging  & $.1$ radians  \\ \hline
\end{tabular}
\end{center}
\end{table}

\clearpage

\section{Supplementary Figures}

\begin{figure*}[ht]
\centering
\includegraphics[width=\columnwidth]{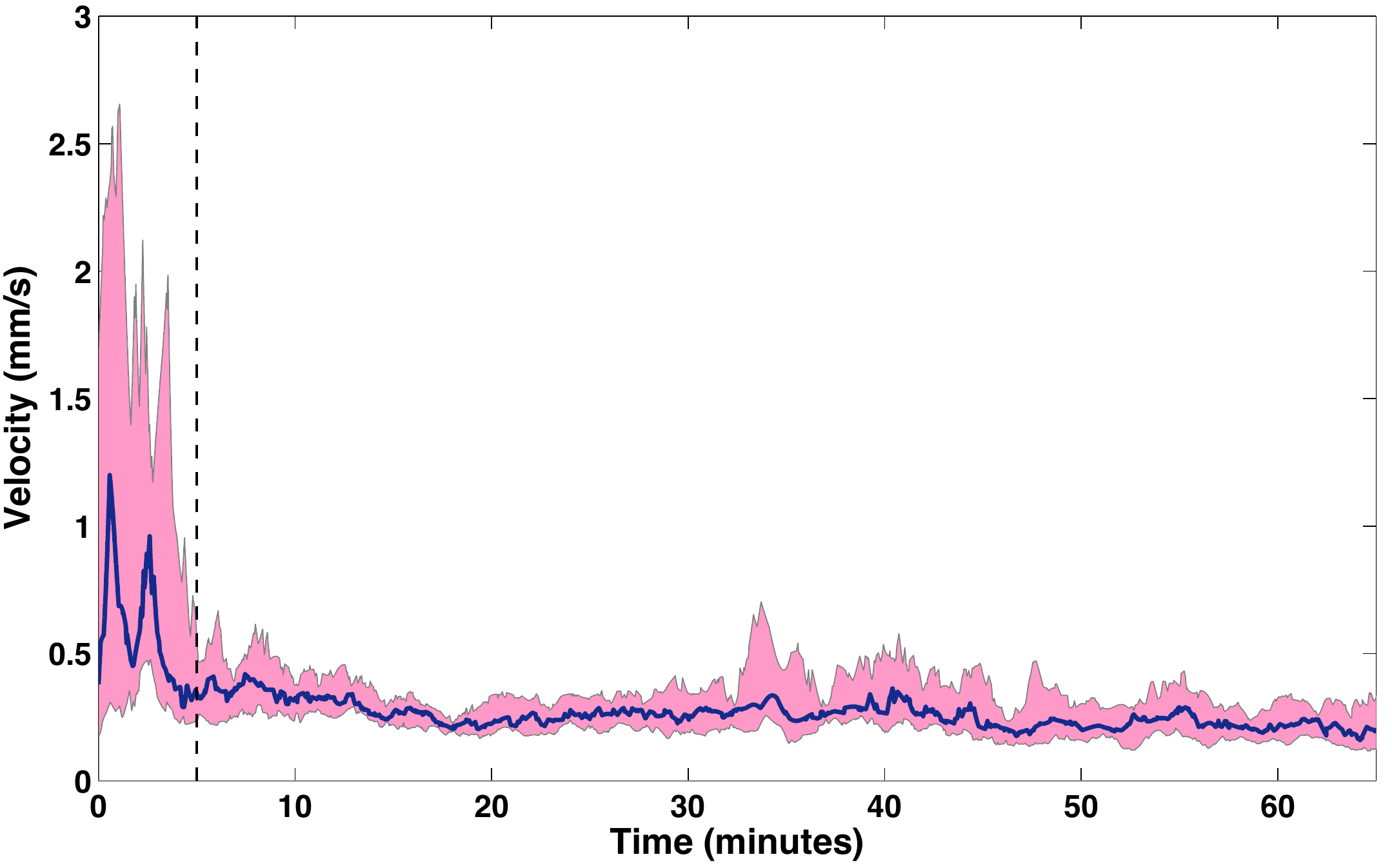}
\caption{Flies adapt to the arena in about five minutes. Median movement velocity for male flies ($N=59$) during the 65 minute filming period.  Pink regions represent the 25\% and 75\% quantiles at each point in time.  Note that almost all adaptation occurs within the first five minutes placed in the dish.  This region is excluded from our analyses}\label{S1}
\end{figure*}

\begin{figure*}[ht]
\centering
\includegraphics[width=.5\columnwidth]{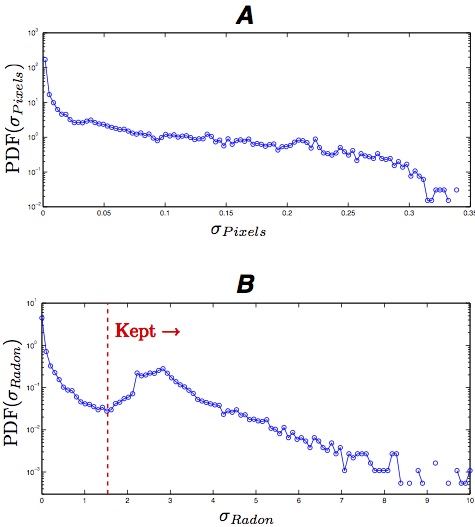}
\caption{Radon vs. pixel representation of images.  A) Probability density function of pixel standard deviations.  B) Probability density function of Radon pixel standard deviations.  Note the clear minimum that exists in B), allowing for an effective reduction in the number of pixels necessary to represent the data.}\label{S2}
\end{figure*}

\begin{figure*}[ht]
\centering
\includegraphics[width=\columnwidth]{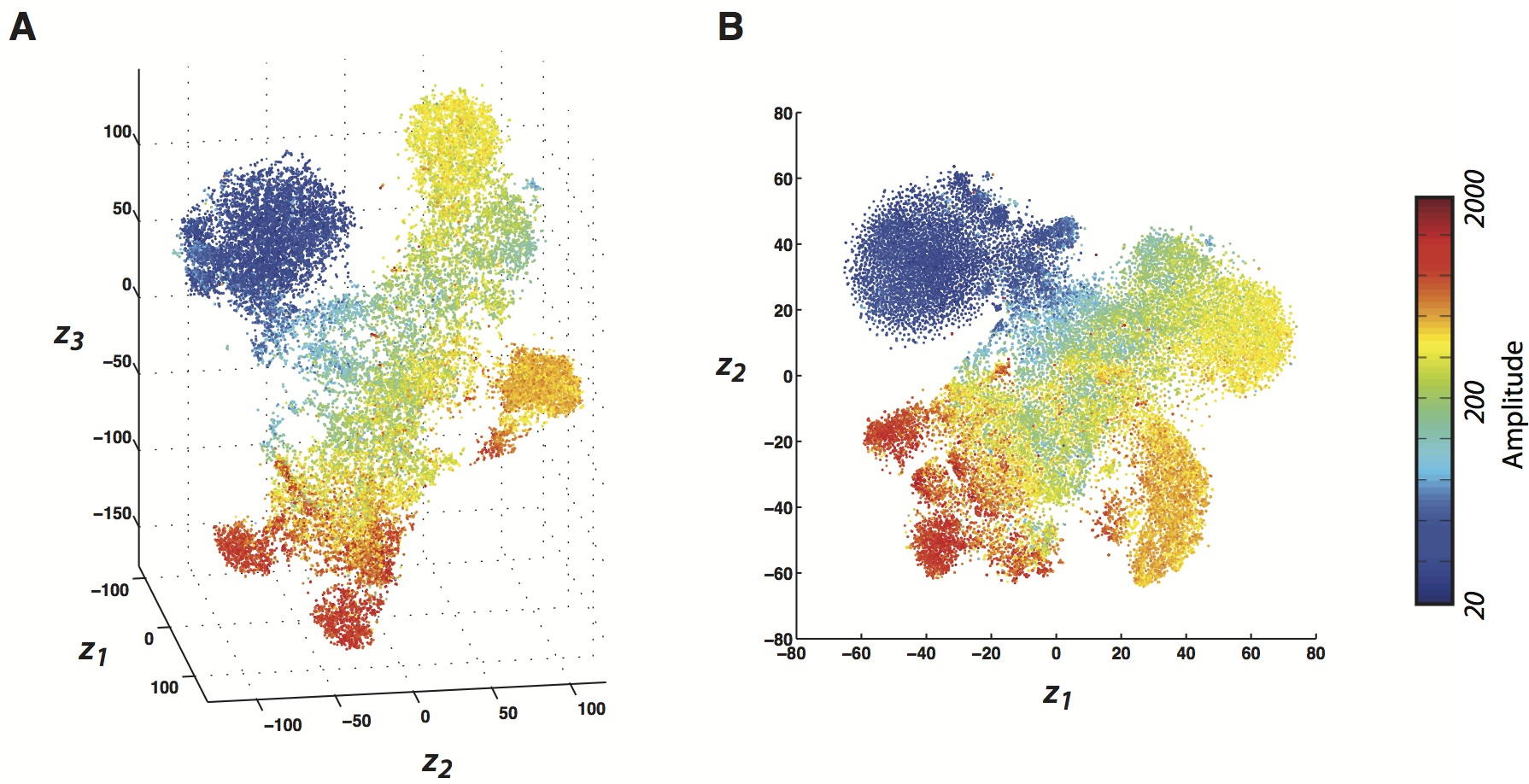}
\caption{Comparison between embedding into 3-D (A) versus 2-D (B) via t-SNE.  Other than the embedding dimension, all other parameters remain constant.  Color labels are proportional to the logarithm of the normalizing amplitude ($\sum_{k,f} S(k,f;t)$) for each point in the training set.  There is a 2\% improvement in the error function (Equation D1) in the 3-D case as compared to the 2-D embedding (2.9 versus 3.3 bits out of a total of 20.6 bits for the transition matrix $P$).}\label{S3}
\end{figure*}

\begin{figure*}[ht]
\centering
\includegraphics[width=\columnwidth]{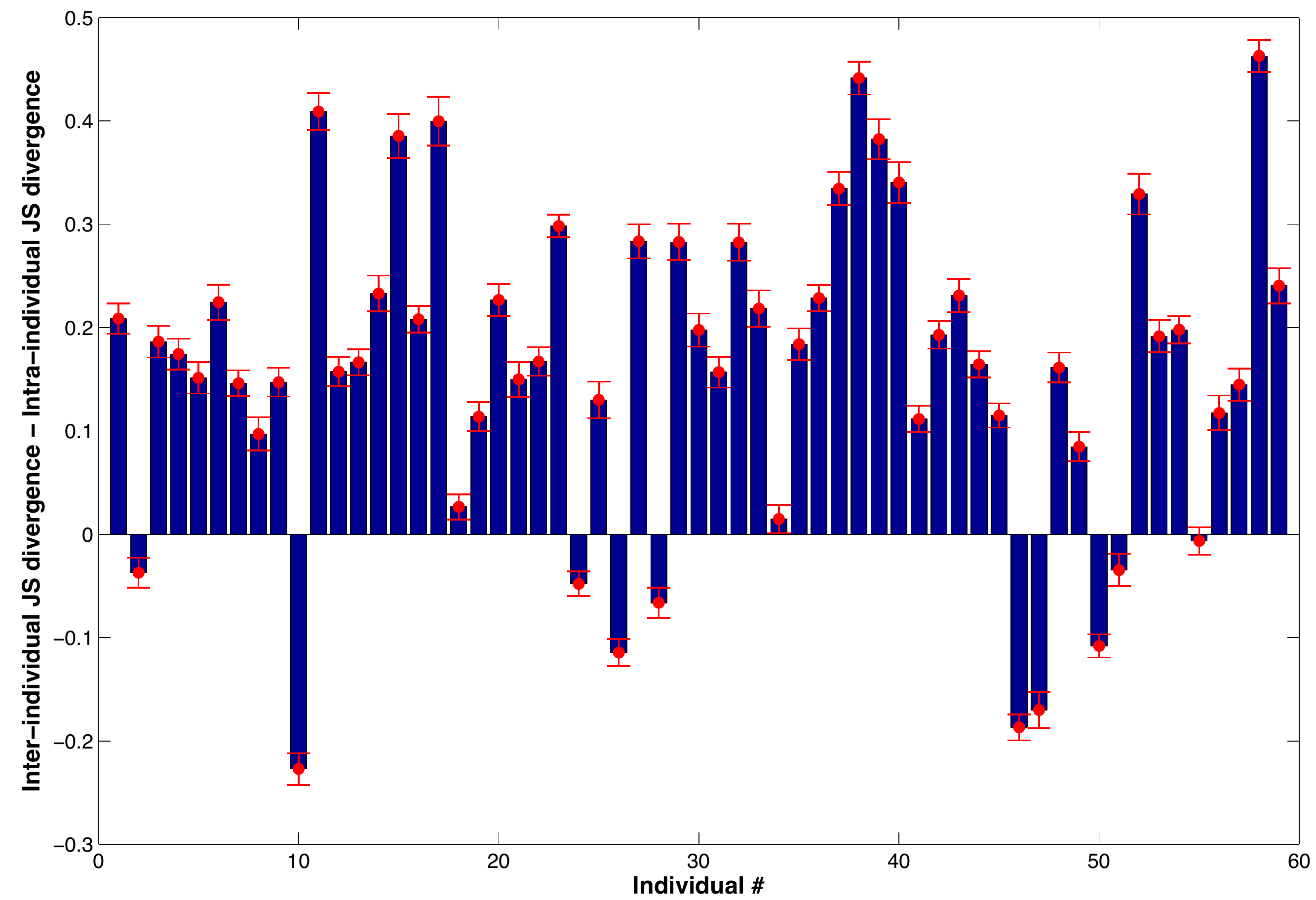}
\caption{Inter- versus intra- individual variation.  Plotted is the inter-individual variation minus the intra-individual variation for each of the 59 male flies.  Here, we measure the difference between two behavioural spaces as the Jensen-Shannon (JS) divergence between their respective probability densities.  Intra-individual variation is measured as the JS divergence between maps generated from the first 20 minutes and the last 20 minutes of an individual data set.  Inter-individual variation is measured as the median JS divergence between the behavioural space between individuals.  Error bars are calculated via N-1 bootstrapping on the inter-individual variance.  Accordingly, a positive value on this plot implies that the two portions of the data set are more similar to each other than they are to other individuals' behavioural spaces.  We find that 48 of the 59 individuals display significantly less intra- than inter-individual variations ($\mu = .16 \pm .02$).}\label{S4}
\end{figure*}

\begin{figure*}[ht]
\centering
\includegraphics[width=\columnwidth]{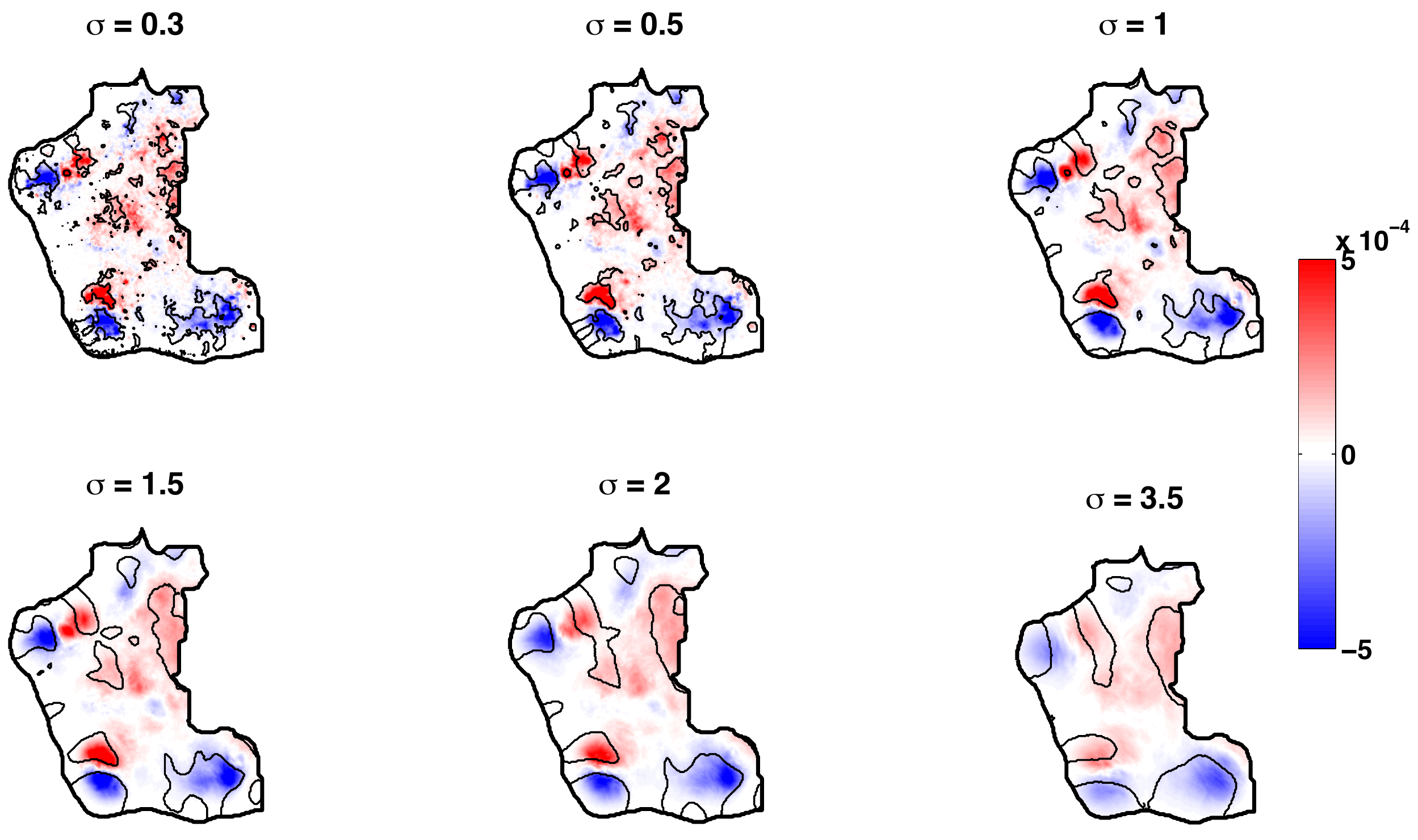}
\caption{Effect of varying the smoothing parameter, $\sigma$, on the identified male/female wing movement distinctions.  Plots are of the difference between the median female region-normalized probability density function and the median male region-normalized probability density function for various values of $\sigma$.  Lines within the regions encircle regions where the \emph{p}-Value of the Wilcoxon rank sum test are less than .01. }\label{S5}
\end{figure*}

\begin{figure*}[ht]
\centering
\includegraphics[width=\columnwidth]{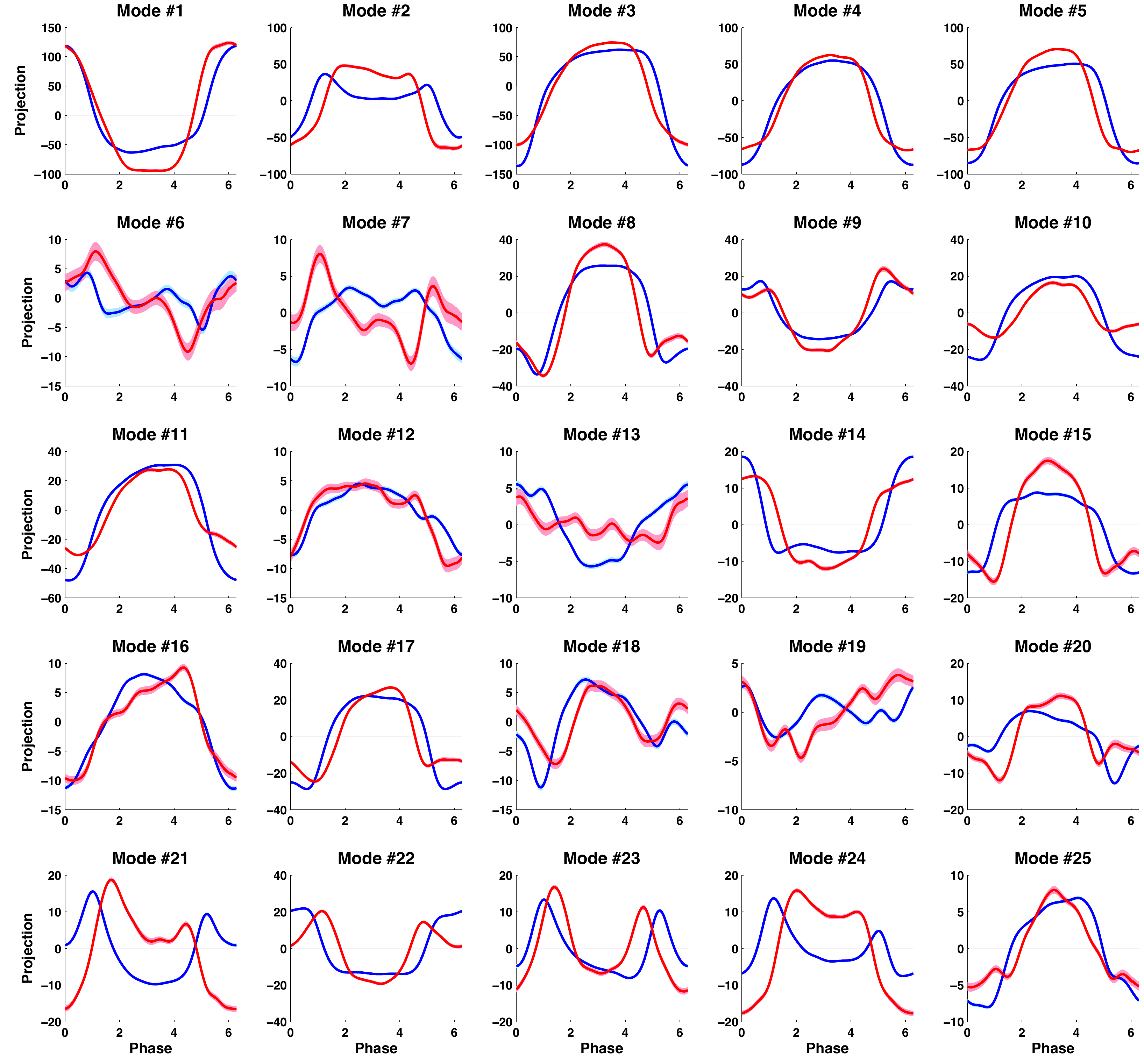}
\caption{Postural space periodic orbits for left wing grooming behaviours in Fig 11 in the main text.  Blue curves are the average orbits for region $(i)$, and red curves are average orbits for region $(ii)$.  Line thicknesses represent that standard error of the mean at each phase.}\label{S6}
\end{figure*}

\clearpage
\section{Supplementary Movie Legends}

\emph{\textbf{Supplementary movies can be obtained by emailing GJB (gberman@princeton.edu)}}
\vspace{.1in}

\noindent {\bf Movie S1.}
Raw video data of a behaving fly (left) and the corresponding segmented and aligned data (right).

\noindent {\bf Movie S2.}
Dynamics in behavioural space. Raw video of a behaving \emph{D. melanogaster} (middle) is displayed alongside coordinates of the fly's position within the filming apparatus (left) and its position in the embedded behavioural space (right).  The red circles represent the positions in the appropriate coordinate system and the trailing lines are the positions traversed in the previous .5~s.  The light blue shading indicates that a particular behaviour is being performed, and the blue text below the video of the fly gives a coarse label for the behaviour.   The first portion of the movie is 5~s, played at real time (indicated by ``Real Time" above the fly video), and the subsequent portion of the movie is slowed down by a factor of 5 for clarity (indicated by ``Slowed $5\times$").

\noindent {\bf Movies S3-11.}
Each movie is a mosaic of multiple instances of specific regions in behavioural space as displayed in Fig \ref{movie_region} and Table \ref{movietable}. Every movie contains multiple segments from many different individuals and are slowed by a factor of 4 for clarity.  

\begin{table}[h]
\caption{Behavioural Movies \label{movietable}}
\begin{center}
\begin{tabular}{l|l}
\textbf{Movie} & \textbf{Label} \\
\hline
Movie S3 & Idle \\
Movie S4 & Right wing grooming \\
Movie S5 & Left wing grooming \\
Movie S6 & Left wing and legs grooming \\
Movie S7 & Wing waggle \\
Movie S8 & Abdomen grooming \\
Movie S9 & Running \\
Movie S10 & Front leg grooming\\
Movie S11 & Head grooming\\
\end{tabular}
\end{center}
\end{table}

\noindent {\bf Movie S12.}
Composite movie (slowed by a factor of 4) of randomly chosen instances of flies from the male-preferred behavioural region in Fig \ref{male_female_figure} of the main text.

\noindent {\bf Movie S13.}
Composite movie (slowed by a factor of 4) of randomly chosen instances of flies from the female-preferred behavioural region in Fig \ref{male_female_figure} of the main text.

%\bibliography{Berman_ScienceNew_AllRefs}

\end{document}